Kseniia Mamaeva †, Hodjat Haijan †, Carolyn Elliott, Hannah Killeen, Teodora Faraone, Larisa Florea, Colm Delaney, and A. Louise Bradley*

# Integration of quantum dots at the tips of single plasmonic bipyramid nanoantennas for strong coupling at room temperature

**Abstract:** Achieving strong coupling between excitons of colloidal semiconductor quantum dots (QDs) and localized surface plasmon polaritons (LSPs) is critical for advanced room-temperature quantum emitter and sensing applications. A key challenge is to have precise control of the emitters position with respect to an individual plasmonic nanostructure. Here, we present room temperature strong coupling between QDs and a single gold nano-bipyramid (BPs). The selection of the bipyramid plasmonic nanocavity offers access to a single hotspot with a very small mode volume. The localization of QDs at a single hotspot is achieved *via* plasmon-triggered two-photon polymerization. This technique exploits the enhanced electric field at the BP tip to selectively polymerize a photosensitive QD-containing formulation. Room-temperature scattering spectra of a 3-QD-BP system reveal Rabi splitting of 349.3 meV and a coupling strength of 175.68 meV. The with distinct anti-crossing behavior is confirmed by simulations. This approach simplifies QD integration for strong coupling systems compared to previous methods. These results indicate a scalable platform for solid-state quantum technologies with colloidal QDs, enabling exploration of exciton-plasmon interactions and further advancement of applications in quantum optics and quantum sensing under ambient conditions.

## 1 Introduction

Solid-state quantum emitters, such as epitaxial and colloidal quantum dots (QDs), are promising for a wide range of optoelectronic and quantum technologies. Colloidal QDs offer high quantum yields, emission tunability and compatibility with on-chip integration [1]-[3]. Control over the light matter-interaction is extensive area of research for strong coupling with the optical modes, emission control [4]-[6] and nonlinear interactions [2], [7], [8]. Embedding QDs within nanocavities [9], [10], photonic crystals [11]-[15], or plasmonic structures [16]-[20] creates the confined electromagnetic environment necessary to achieve these effects but the efficiency is highly dependent on QD placement [21], [22].

Various techniques have been explored for the deterministic placement of QDs [23], [24]. Epitaxial QDs can be positioned within nanophotonic structures by directing their growth into lithographically defined cavities [25]-[28]. For instance, Badolato et al. aligned QDs with photonic crystal cavities for strong coupling [25], while other works achieved integration into micropillars or Bragg gratings [26]-[28]. Alternatively, post-growth embedding into dielectric cavities has been demonstrated to enhance light-matter interaction [29]. However, these systems typically operate at cryogenic temperatures and are challenging to integrate with plasmonic structures due to fabrication incompatibilities. In contrast, colloidal core–shell QDs can operate at room temperature and offer greater flexibility. One deterministic approach involves atomic force microscopy (AFM)-based nanomanipulation. Other approaches include positioning a single gold nanoparticle near a CdSe/ZnS QD [30], dip-pen nanolithography using a QD-loaded AFM tip to deposit QDs at predefined sites [31], transferring nanodiamonds with nitrogen-vacancy centers onto plasmonic antennas [32], or assembling tunable cavities above monolayers of colloidal nanocrystals [22]. Alternatively, researchers are exploring the use of directed driving forces for emitter placement. Techniques include capillarity-driven assembly into hollow apertures [33]-[35], and plasmonic or optical tweezers, which exploit gradient optical forces and polarization control to trap and position nanoemitters near plasmonic structures with high spatial precision [36].

In this manuscript, we explore plasmon-assisted two-photon polymerization for the deterministic placement of quantum emitters at the tip of single gold BPs, creating a platform for

Kseniia Mamaeva and Hodjat Haijan contributed equally to this work.
*Corresponding author: **A. Louise Bradley**, School of Physics, Trinity College Dublin, Dublin, Ireland and IPIC, Tyndall National Institute, Cork, Ireland; e-mail: bradlel@tcd.ie; https://orcid.org/0000-0002-9399-8628
**Kseniia Mamaeva:** School of Physics, Trinity College Dublin, Dublin, Ireland and IPIC, Tyndall National Institute, Cork, Ireland; https://orcid.org/0000-0003-3232-3974

**Hodjat Haijan:** School of Physics, Trinity College Dublin, Dublin, Ireland; https://orcid.org/0000-0001-6564-6273
**Carolyn Elliott:** School of Physics, Trinity College Dublin, Dublin, Ireland and IPIC, Tyndall National Institute, Cork, Ireland
**Hannah Killeen:** School of Physics, Trinity College Dublin, Dublin, Ireland
**Teodora Faraone, Larisa Florea and Colm Delaney:** School of Chemistry, Trinity College Dublin, Dublin, Ireland

investigating and achieving strong coupling between excitons and localized surface plasmons. At the nanoscale, light–matter interaction can enter the strong coupling regime, where coherent energy exchange between the plasmonic mode and the excitonic state dominates over their respective loss rates, leading to the formation of hybrid polaritonic states [37]-[39]. In the weak coupling regime, the emission rates of the quantum emitter can be boosted via the Purcell effect [9], [40]-[44]. On the contrary, if the coupling strength ($g$) surpasses the total loss rate, the hybrid emitter-cavity system can operate in the strong coupling regime as seen with monolayer semiconductors [12], [16], [45]-[50], single molecule [17]-[19], and QDs [10], [13]-[15], [20], [34], [51]-[53]. In this case, the hybrid polaritonic quasiparticles are created due to the coherent energy exchange between the coupling subsystems, occurring on time scales much faster than their dissipative dynamic processes.

Room temperature strong coupling can be primarily obtained by the integration of N molecules and their J-aggregates, as well as two dimensional transition-metal dichalcogenides (TMDCs), to metasurfaces [12], [45]-[50] and plasmonic nanostructures [16]-[20]. In addition to operating at room temperature, plasmonic systems offer several advantages, including an ultra-small mode volume ($V_m$), a widely and geometrically tunable bandwidth, and efficient integration with quantum emitters. Achieving room-temperature strong coupling between quantum emitters and plasmonic nanostructures with deep-subwavelength mode confinement is a key requirement for scalable on-chip quantum photonic technologies [10], [21], [34], [51]-[53]. Room temperature Rabi splitting of 176 meV was reported with silver bow-tie plasmonic cavities, fabricated by the electron beam lithography (EBL), with the gap sizes around 20 nm and loaded with a single QD [34]. In a similar approach, room temperature strong coupling of a mesoscopic colloidal QD to an EBL-fabricated plasmonic nanoresonator at the apex of a scanning probe exhibited Rabi splitting of up to 110 meV [22]. Rabi splitting of up to 200 meV can be achieved in these so-called nanoparticle-on-mirror (NPoM) configurations by positioning a single QD in the gap between a bottom reflector and a scanning plasmonic antenna tip [52] or a plasmonic nanoparticle [53]. However, the presence of the plasmonic bottom reflector may hinder the practical applications regarding the on-chip integration of NPoMs as well as the out-of-plane two-port functionality of a transmissive device. To address this issue, it was recently reported that placing a core−shell QD beneath one end of an angled Au nanorod on a transparent substrate – forming what is known as a wedge nanogap cavity (WNC) configuration – yields a large Rabi splitting of 234 meV due to the strong localization of the electric field within the QD's nanoshell [21]. Although this approach is very effective, it requires considerable trial and error to (i) obtain a single QD using the self-assembly method with the linker molecule 11-Amino-1-undecanethiol hydrochloride and (ii) precisely position the QD beneath the Au nanorod, making the feasibility and reproducibility of the WNC configuration challenging. It has been recently reported that by controlling the local distribution of QDs in a photosensitive formulation through a plasmon-triggered two-photon polymerization (P-2PP) technique it is possible to hybridize cubic plasmonic nanoparticles and QDs to achieve enhanced quantum emission and photoluminescence (PL) [41].

In this study, we demonstrate the precise positioning of colloidal semiconductor QDs at the tips of plasmonic gold nano-bipyramids (BPs) to achieve strong coupling between excitons and localized surface plasmon polariton (LSP) modes at room temperature. Plasmonic gold bipyramid nanoparticles (BPs) are highly promising for applications in strong coupling due to their sharp tips and widely tunable LSP resonances. The intense localized fields at their tips facilitate the formation of hybridized plasmon-exciton states once BPs are coupled to emitters such as TMDC [16], [17]. Herein, a single Au BP coupling with 3 QDs at a single BP tip is achieved using plasmon-assisted two-photon polymerization (P-2PP) technique [41], [54]-[57]. The overlap between the significantly enhanced electric field at the BP tip and a photosensitive formulation containing diluted QD concentrations provides a method to localize the QDs at the bipyramid tip where the maximum field enhancement occurs. The hybrid system's measured scattering spectra — supported by simulations and semi-analytical calculations — reveal a large Rabi splitting and anti-crossing behavior. This confirms access to the strong coupling regime. Our approach of combining the bipyramid plasmonic nanoparticle geometry and plasmon-triggered (P-2PP) offers a reproducible and straightforward method for integrating QDs with single plasmonic nanoantennas

## 2 Results and discussion

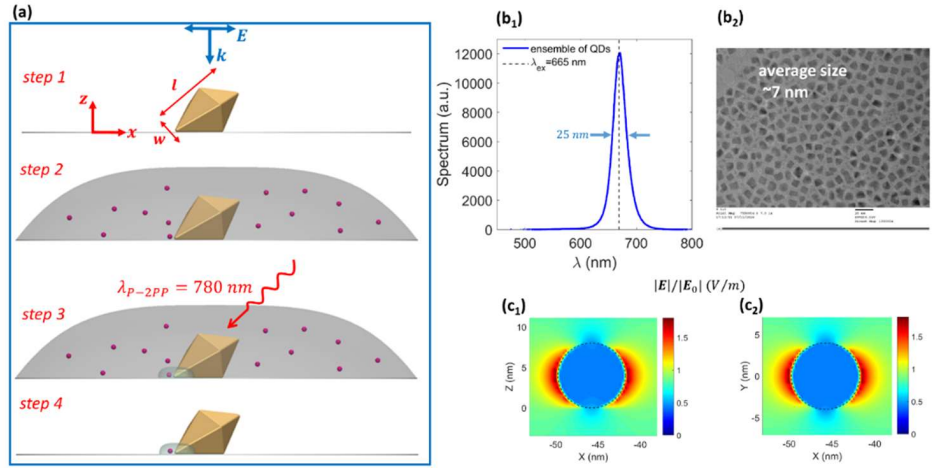

**Figure 1.** (a) Schematics showing the steps of preparation the hybrid QD-BPs based on the P-2PP approach. The length and width of the BP are labeled as $l$ and $w$, respectively, with an aspect ratio of $AR = l/w$. The x-polarized incident plane wave in the simulations is also shown

schematically. (b1) The measured photoluminescence (PL) spectrum with $FWHM = 25\ nm$. (b2) TEM image of the colloidal CdSeS/ZnS alloyed QDs on a lacey carbon film supported by a 300 mesh copper TEM grid with the average diameter of ~7 nm. The vertical dashed line in (b1) indicates $\lambda_{ex} = 665\ nm$ in used in the model, close to the peak of the measured PL spectrum at 669 nm, while the simulated scattering spectrum of the QD is maximum at $\lambda_{ex,s} = 640\ nm$. Detailed analysis of the optical properties of the QD are provided in Fig. S2. (c1) and (c2) show simulated electric field maps for a CdSeS/ZnS QD ($r = 4\ nm$) on the SiO2 substrate under the incident plane wave at $\lambda_{ex}$.

The steps of the sample preparation are schematically shown in Figure 1(a). Gold bipyramids (BPs) were synthesized using a seed-mediated growth method [58], [59], starting with penta-twinned gold seeds prepared by reducing HAuCl$_4$ with sodium borohydride. These seeds were introduced into a growth solution containing gold and silver precursors, CTAB, hydrochloric acid, and ascorbic acid. Reaction conditions were optimized to control the bipyramids' size, aspect ratio, and tip sharpness, influencing their plasmon resonance. The prepared BPs were deposited on glass slides via spin coating (step 1). The CdSeS/ZnS alloyed QDs (Sigma-Aldrich) are dispersed in a polymerizable solution consisting of 99 wt% TMPET (Trimethylolpropane ethoxylate triacrylate) and 1 wt% PBPO (phenylbis (2,4,6-trimethylbenzoyl) phosphine oxide), then drop-cast onto the BPs on a glass slide. A solution with quantum dots (QDs) was applied (step 2), and the P-2PP technique was employed to localize QDs near the BP tips. This nonlinear optical process selectively polymerizes the polymer at regions of enhanced electric fields created by the BPs (step 3). Finally, the sample is cleaned with only the QDs confined to the polymerized material at the BP tips remaining (step 4). Details are provided in the Methods.

The measured PL spectrum of an ensemble of QDs, with an excitonic resonance at $\lambda_{ex} = 665\ nm$ and a full width at half maximum (FWHM) of 25 nm, is shown in Figure 1(b), accompanied by a transmission electron microscopy (TEM) image of the ensemble in Figure 1(c), which indicates an average QD diameter of ~7±1 nm. Based on the measured QD diameter distribution, shown in the Supporting Information Figure S2.6, we modeled the QD in the finite-difference time-domain (FDTD) simulations [60] as a sphere with a radius of $r = 4\ nm$ and a Lorentzian dielectric permittivity. The simulated scattering spectrum of the QD on a SiO$_2$ substrate indicates a peak at $\lambda_{ex,s} = 640\ nm$ with a half-linewidth of $\gamma_{ex,s} = 106.34\ meV$, Figure S2, and results in a weak excitonic enhancement of the electric field when an x-polarized plane wave is normally incident on the QD, as illustrated in Figs. 1(c1) and 1(c2). Note that the electric field at the surface of the bare substrate is $|E_0| \approx 0.83$. Therefore, the mode profiles shown here also represent the electric field enhancement, $|E|/|E_0|$.

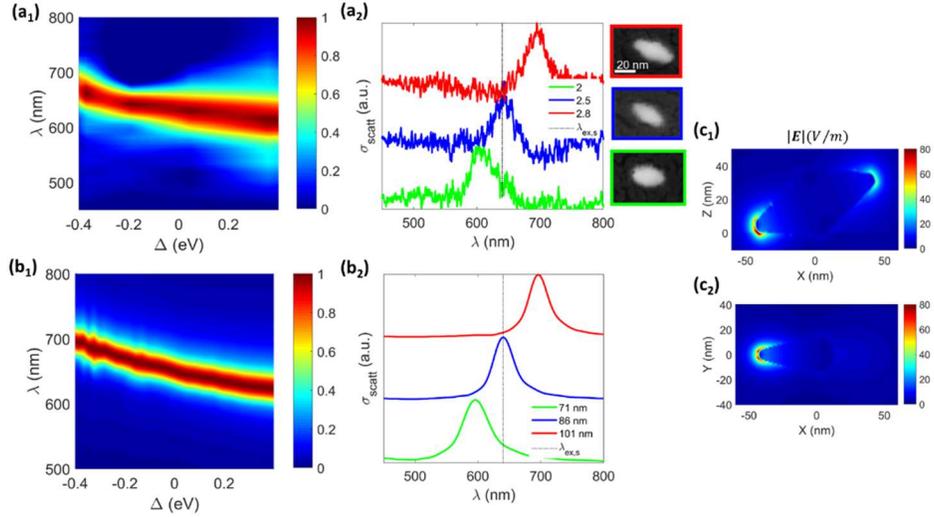

**Figure 2**. (a1)-(a2) Experimental scattering spectra of the bare Au BPs of varying lengths on the SiO2 substrate, corresponding to aspect ratio between 2 to 2.5. (b1) and (b2) show the corresponding normalized simulated scattering spectra of the bare BPs with varying lengths. ($71\ nm < l < 101\ nm$) and constant values of the width and the radius of the BP tip [$w = 40\ nm$ and $r_{BP} = 4\ nm$]. The vertical dotted line in (a2) and (b2) indicates the scattering peak wavelength of the QD; i.e. $\lambda_{ex,s} = 640\ nm$. Note that for the solid-blue curve $\gamma_{LSP} = 68.24\ meV$. (c1)-(c2) Side and top-view of the electric field distribution of the LSP of the bare BP on SiO2 for the case of $l = 86\ nm$ at $\lambda_{ex,s}$. The substrate is located at $z < 0$. For the top-view profile, the monitor is located at $z = 4\ nm$.

The measured single bare BP scattering spectra are shown in panels (a1) and (a2) of Figure 2. Plotted as a function of the detuning energy [$\Delta = \hbar(\omega_{LSP} - \omega_{ex,s})$], it is seen that the scattering peak ($\lambda_{LSP}$) is tuned from 600 nm to 700 nm, where $\omega_{LSP} = 2\pi c/\lambda_{LSP}$, $\omega_{ex,s} = 2\pi c/\lambda_{ex,s}$, and $\lambda_{ex,s} = 640\ nm$ is the peak wavelength of the scatering spectrum of the bare QD [see Fig. S2(c)]. As illustrated in Figure 2(a2), along with the scanning electron microscopy (SEM) images of the bare BPs spin-coated onto a glass substrate and imaged with with a thin (~5nm) gold layer, here the aspect ratio of the bare BPs is varied from 2 to 2.8.

The FDTD simulated single BP scattering spectra are shown in panels (b1) and (b2) of Figure 2, as a function of $\Delta$ and the BP length, respectively. Here, the simulated BP dimensions are $w = 40\ nm$, BP tip radius $r_{BP} = 4\ nm$ with BP length in the range $71\ nm < l < 101\ nm$. Note that the fabricated BPs are lithography-free, meaning we do not have precise control over the tuning of their length, width, and tip radius. However, based on the

BP geometrical parameters extracted from the SEM images, the values of $l$, $w$, and $r_{BP}$ used in the simulations were selected to achieve optimal agreement between the measured and simulated spectra. Here $w$ and $r_{BP}$ have been chosen in a way that the BP illustrates a $\sigma_{scatt}$ peak at $\lambda_{ex,s}$ for $l = 86\ nm$ with the half-linewidth of $\gamma_{LSP} = 68.24\ meV$ [$\Delta = 0$ in Figure 2(b1) and the solid blue curve in Figure 2(b2)]. For this case, the side- and top-views of the plasmonic $|E|$ distributions are illustrated in panels (c1) and (c2) of Figure 2, respectively. It is observed that the electric field is 80 times enhanced at the tip of the BP due to the excitation LSP polariton with the excitation wavelength corresponding to $\lambda_{ex,s}$. Further calculations show this resonance, accompanied by the mentioned field enhancement, leads to the mode volume of $V_m = 4451\ nm^3$. Refer to Section 2.2 of the Supplementary Information for details of calculation. Note that here the permittivity data for Au were taken from Reference [61].

It is observed from the results shown in Figure 2 that the spectral location of the resonances of the BPs in the scattering spectra follows the commonly used analytical formulation of the plasmonic nanoparticle $\lambda_{res} = \lambda_0 (l/l_0)^\alpha$ where $\lambda_0$ is a reference resonance wavelength for a given size, $l_0$ is the corresponding length to $\lambda_0$, and $\alpha$ is an empirical scaling factor that depends on the nanoparticle geometry. As investigated in Figure S3, this trend is also observed in the absorption spectrum ($\sigma_{abs}$), leading to a similar feature in the extinction spectrum ($\sigma_{ext}$) as well, where $\sigma_{ext} = \sigma_{scatt} + \sigma_{abs}$e.

The strong field enhancement of the BP together with the small value of $V_m$ suggests that, when a single QD is integrated at the tip of the BP, strong coupling between the BP LSP and the QD exciton is likely to occur, creating two exciton-polariton states with the following eigen energies [49]

$$E_{1,2} = \frac{1}{2}[(E_{ex,s} + E_{LSP}) + i(\gamma_{ex,s} + \gamma_{LSP})] \pm \sqrt{Ng^2 + \frac{1}{4}[\Delta + i(\gamma_{ex} - \gamma_{LSP})]^2} \tag{1}$$

Here, $E_{LSP} = \hbar\omega_{LSP}$ and $E_{ex,s} = \hbar\omega_{ex,s}$ are the energies for the uncoupled BP LSP and the QD exciton scattering resonances, respectively, $N$ is the number of the QDs integrated at the BP tip, and $E_{1,2} = \hbar 2\pi c/\lambda_{1,2}$. If $\Delta = 0$, the coupling strength g can be calculated using Equation 1

$$g_{s,e} = \sqrt{(\hbar\Omega_{s,e})^2 + (\gamma_{ex,s} - \gamma_{LSP})^2}/2 \tag{2}$$

where 's' and 'e' denote values extracted from simulated and experimental scattering spectra, respectively. The schematic of this single QD-BP (1-QD-BP) system (i.e. $N = 1$) and the corresponding simulated scattering spectrum (solid-blue curve) are shown in Figure 3(a). In agreement with Figure S2 and Figure 2(b2), scattering of the QD (solid-red) and the bare BP (dashed-black) are also shown as the references in this figure. In this case, due to the strong coupling between the QD exciton and the BP LSP, two peaks appear at $\lambda_{s1} = 626\ nm$ and $\lambda_{s2} = 684\ nm$ in the simulated σ_scatt spectrum, resulting in a Rabi splitting of $\hbar\Omega_s = 168\ meV$. The coupling strength for this 1-QD-BP system is calculated to be $g_s = 92\ meV$. As an alternative approach, the position-dependent coupling strength can be also numerically calculated based on the QD's dipole moment ($\mu_{QD}$) and normalized electric field at the location of the QD ($E(r_{QD})/|E(r_{QD})|_{max}$) [19], [21]

$$g_s = \frac{E_{ex}}{\sqrt{2\epsilon_0 E_{LSP} V_{eff}}} \mu_{QD} \cdot \frac{E(r_{QD})}{|E(r_{QD})|_{max}} \tag{3}$$

It is observed that the calculated value of $g_s$ based on Equation 3 closely matches the value obtained from Equation (2) for $\mu_{QD} = 0.68\ e.nm$ [19].

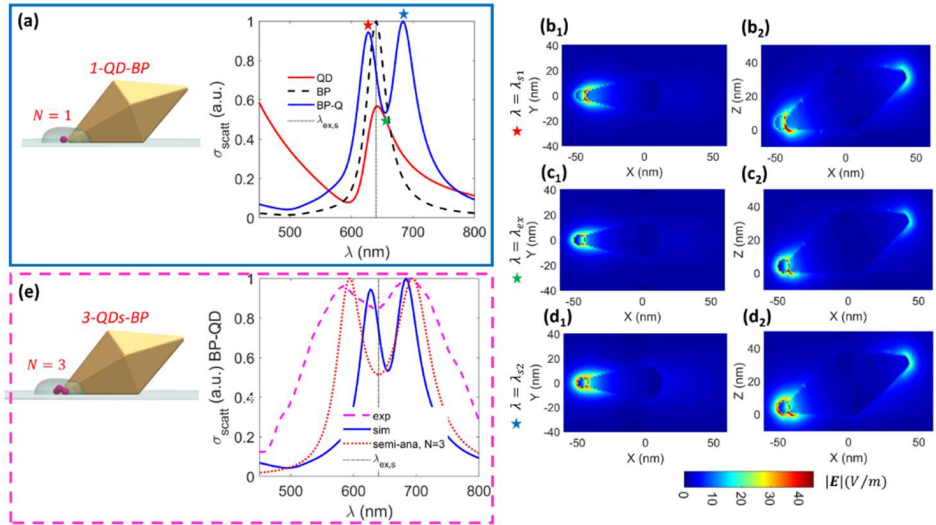

**Figure 3**. (a) Simulated scattering spectra of the bare QD (solid-red), the bare BP (dashed-black) and the 1-QD-BP (solid-blue) on a SiO2 substrate for the case of $l = 86\ nm$. For this curve, the peaks are located at $\lambda_{s1} = 626.8\ nm$ (red star) and $\lambda_{s2} = 683.8\ nm$ (blue star), corresponding to a Rabi splitting of $\hbar\Omega_s = 168\ meV$. (b1), (c1), and (d1) [(b2), (c2), and (d2)] exhibit top-view [side-view] of the $|E|$ distributions at the peaks and the dip of the solid-blue curve in panel (a), respectively. (e) The dashed-pink curve shows the experimental scattering spectrum of the QD-BP

for the case of $l = 86\,nm$, showing two peaks located at $\lambda_{e1} = 580.5\,nm$ and $\lambda_{e2} = 693.9\,nm$, equivalent to a Rabi splitting of $\hbar\Omega_e = 349.3\,meV$. Considering $N = 3$ in Eq. (3), the semi-analytical dashed-red curve closely fits the dashed-pink one at $\lambda_{e1}$ and $\lambda_{e3}$, indicating that three QDs are attached to the tip of the BP in the experiment. The vertical-dotted line in panels (a) and (e) indicate $\lambda_{ex,s}$.

For the peaks and the dip of the solid blue curve in Figure 3(a), marked by the red, blue, and green stars, the top-view [(b1), (c1), and (d1)] and side-view [(b2), (c2), and (d2)] $|E|$ distributions are illustrated in Figure 3 as well. It is observed that the mode profiles inherit both excitonic and plasmonic features of those shown in Figures 1(c1)–1(c2) and Figures 2(c1)–2(c2), indicating the exciton-polariton characteristics of the new hybrid modes supported at $\lambda_{s1}$ and $\lambda_{s2}$.

The dashed pink curve in Figure 3(e) represents the experimental scattering spectrum of the synthetized QD-BP sample, showing two peaks at $\lambda_{e1} = 580.5\,nm$ and $\lambda_{e2} = 693.9\,nm$, corresponding to a large Rabi splitting of $\hbar\Omega_e = 349.3\,meV$. This value is noticeably larger than the one obtained from the simulated 1-QD-BP system [solid blue curve in Figure 3(e)], suggesting that more than a single QD might be integrated at the tip of the BP in the experiment. We compare the experimental scattering spectrum with the one obtained using the following semi-analytical relation [19], [21]

$$\sigma_{scatt}(\omega) \propto -Im[\hbar\omega - (E_{LSP} + i\hbar\gamma_{LSP}) - \frac{Ng_s^2}{\hbar\omega - (E_{ex,s} + i\hbar\gamma_{ex,s})}]^{-1} \quad (4)$$

Taking $N = 3$ into account, as expected from the QD concentration, the semi-analytical result obtained using Equation 4 is shown by the dotted red curve in Figure 3(e). As seen, the semi-analytical approach reproduces the spectral locations of the peaks of the experimental $\sigma_{scatt}$, i.e., $\lambda_{e1,e2}$. Furthermore, by substituting $\hbar\Omega_e$ into Equation 2, the calculated value of the experimental coupling strength is $g_e = 175.68\,meV$. In other words, $g_e \cong \sqrt{N}g_s$ where $N = 3$, confirming that 3 QDs are attached to the BP tip in the experiment. We refer to this system as the 3-QDs-BP case, which is schematically shown in Figure 3(e). This is also supported by the fact that, according to the experimentally measured concentration of QDs (0.0385mg/mL) from the TEM data, the mode volume includes approximately $\overline{N} = 3 - 4$ QDs, as discussed in Section 2.5 of the Supporting Information. Consequently $g_1 = g_e/\sqrt{3} = 102\,meV$ can be considered as an experimental value of the coupling strength of the 1-QD-BP in this study and is in agreement with other previously reported for bow-tie nano-antenna [34], NPoM [10] or WNC configuration [21]. However, this approach 1) only requires single plasmonic nanoresonator, which can be chemically synthesized and provides sharper tips, and 2) offers relative simplicity of integration of the emitter at the nano-resonator hot-spot. Further investigations show that $g_1$ satisfies the following rules of thumb, providing additional support for the fulfillment of the strong coupling condition at room temperature in this study: $g_1 > |\gamma_{ex} - \gamma_{LSP}|/2$ and $g_1 > \sqrt{\gamma_{ex}^2 + \gamma_{LSP}^2}/2$

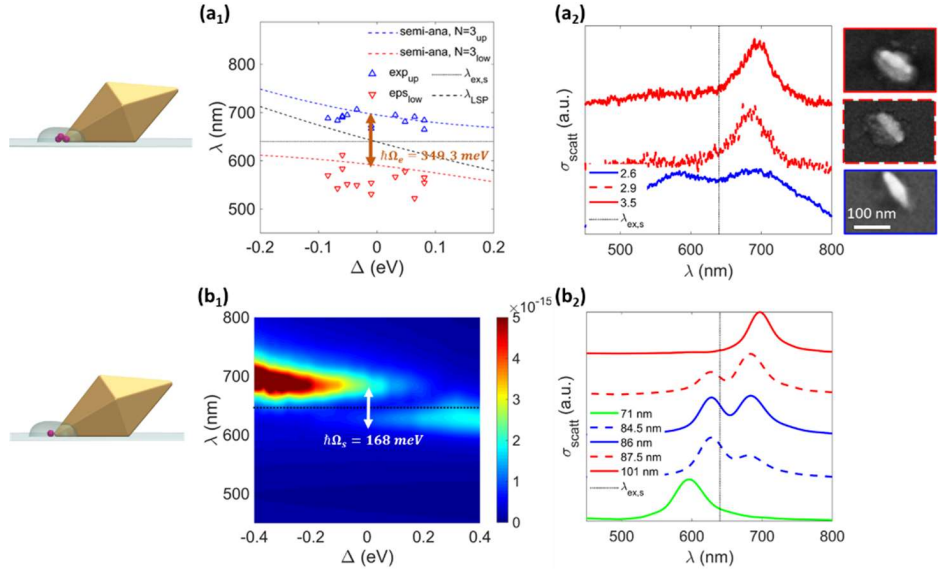

**Figure 4.** (a1) Anti-crossing of the measured scattering spectra indicates achievement of the strong coupling between 3-QDs excitons and the BP LSP, evidenced by the Rabi splitting of the upper and lower hybrid exciton-plasmon modes, i. e. $\hbar\Omega_e = 349.3\,meV$. The dashed-blue and dashed-red curves are obtained semi-analytically from Eq. (1) for $N = 3$. The dotted-black line and the dashed-black curve show the dispersion of the BP LSP and $\lambda_{ex,s}$, respectively. In agreement with (a1), measured spectra of the 3-QDs-BP system for different LSP peak wavelengths, with the corresponding SEM images, are illustrated in (a2). Here, the solid-blue and solid-red curves represent the cases where the QDs are, respectively, strongly and weakly coupled with the BP. (b1) The simulated scattering spectra of the 1-QD-BP system, shown as a color map, clearly exhibits anti-crossing behavior with a Rabi splitting of $168\,meV$. Using $w = 40\,nm$ and $r_{BP} = 4\,nm$, the spectra are obtained by varying the BP length ($71\,nm < l < 101\,nm$). (b2) Simulated scattering spectra are shown for different values of the BP length. Note that the horizontal [(a1) and (b1)] and vertical [(a2) and (b2)] dotted lines indicate $\lambda_{ex,s}$. Schematics of the 3-QDs-BP and 1-QD-BP cases are shown on the left-hand side, as references.

Figure 3 provides further evidence for the strong coupling conditions in the 1-QD-BP and 3-QD-BP systems, resulting in the formation of new exciton-polariton states. A key signature of the strong coupling is the observation of the Rabi splitting ($\hbar\Omega_{s,e}$), which manifests as an anti-crossing behavior in the scattering spectrum, as well as the absorption and the extinction spectra. This characteristic serves as a crucial verification of the strong coupling regime in the QD-BP system. To this end, the resonance wavelengths taken from the measured scattering spectra ($\lambda_{e1,e2}$) of the 3-QDs-BP system are presented in Figure 4(a1) marked by the blue and red triangles. Taking $N = 3$ and $g_s$, the dashed blue and dashed red curves represent the corresponding fitted results obtained from Equation 1. The fitted curves are in good agreement with the experimental data, verifying the obtained value of the experimental Rabi splitting $\hbar\Omega_e$ at $\Delta = 0$ for the 3-QD-BP system and indicating the anti-crossing characteristic. Note that $\lambda_{ex,s}$ (dotted-black) and $\lambda_{LSP}$ (dashed-black) are illustrated in Figure 4(a1) as references. Accompanied by the corresponding SEM images and in agreement with Figure 4(a1), the measured scattering spectra for the three different lengths of the BPs are shown in Figure 4(a2). By modifying $l$, the transition of the measured $\sigma_{scatt}$ spectra from the strong (solid-blue) to the weak (solid-red) coupling regime is evident.

Considering $w = 40\ nm$, $r_{BP} = 4\ nm$, and $71\ nm < l < 101\ nm$, simulated scattering spectra of the 1-QD-BP system as a function of wavelength and the detuning energy are shown in Figure 4(b1), exhibiting a clear anti-crossing with the Rabi splitting of $\hbar\Omega_s$ at $\Delta = 0$. As illustrated in Figure S4, this anti-crossing feature is obviously observed in the absorption and extinction spectra as well. This characteristic is clear in Figure 4(b2), where the simulated $\sigma_{scatt}$ spectra are shown for increasing values of $l$.

## 3 Conclusions

In conclusion, we demonstrated the localisation of colloidal semiconductor QDs at one tip of single plasmonic Au BPs using plasmon-assisted TPP. Strong coupling between excitons and LSPs is observed at room temperature. Our approach leverages the significantly enhanced electric field at the BP tip to precisely place the QDs using the P-2PP technique. This method successfully overcomes key challenges in integrating quantum emitters at the hotspot of a plasmonic nanostructure, offering reproducibility, simplicity, and the use of a single plasmonic open nanocavity. The 3-QDs-BP system achieved a room-temperature Rabi splitting of 349.3 meV with a coupling strength of 175.68 meV, verified through simulations and semi-analytical calculations. Furthermore, the simulation of the 1-QD-BP system demonstrates a coupling strength of 102 meV. Overall, our work provides a simple and feasible approach for integrating colloidal QDs with a single nanoresonator. This paves the way for the development of practical solid-state quantum devices that operate at room temperature, with potential applications in quantum sensing and advanced nanophotonic devices.

## 4 Methods

**Gold bipyramid synthesis:** Gold bipyramids were synthesized using a seed-mediated growth technique. For bipyramid growth, the method of Liu et al. [59] was adapted. A growth solution was prepared by combining 0.5 mL of 10 mM HAuCl4, 0.1 mL of 10 mM silver nitrate (AgNO3), and 10 mL of 0.1 M cetyltrimethylammonium bromide (CTAB). The solution was acidified with 0.3 mL of 1 M hydrochloric acid (HCl) to achieve a pH of approximately 3–4. Reduction of $Au^{3+}$ to $Au^+$ was facilitated by adding 0.08 mL of 0.1 M l-ascorbic acid, followed by the introduction of 60 μL of the aged seed solution. This growth reaction was maintained at 30°C in a water bath for 2 hours.

During synthesis, the concentration of reagents such as AgNO3, HCl, and ascorbic acid was systematically varied to produce bipyramids of different sizes, aspect ratios, and tip sharpness. After synthesis, the samples were centrifuged and washed multiple times with millipore water to remove excess surfactants and unreacted reagents. The final dispersions were stored at room temperature in aqueous solutions for further characterization. Further details are provided in the Supporting Information.

**Preparation of the hybrid BP-QD samples:** The CdSeS/ZnS QDs were commercially obtained from Sigma-Aldrich. To prepare the samples for 2PP, BPs were synthesised and spin-coated onto substrates. The spin-coating process was optimized to achieve an appropriate concentration of BPs—ensuring sufficient density for coverage within the patterned regions while avoiding clustering, which could result in overlapping particles or micro-explosions caused by excessive field enhancement.

A polymerizable solution was then applied to the coated slides, consisting of 99 wt% Trimethylolpropane ethoxylate triacrylate (TMPET) as the monomer and 1 wt% Phenylbis(2,4,6-trimethylbenzoyl)phosphine oxide (PBPO) as the photoinitiator. The patterns were written using the Photonic Professional GT2 system by Nanoscribe, with a 780 nm circularly polarised femtosecond laser. A CAD interface was used to control the laser power and focal point with sub-micron precision.

Polymerisation was induced at 0.5–3.5% of the 50 mW maximum laser output (sub-threshold regime), enabled by plasmonic field enhancement. This approach confined polymerisation to areas near the BP tips, where the electric fields were strongest. The resulting structures were washed in warm isopropanol to remove unpolymerized solution, ensuring precise localisation of QDs in the polymerised regions.

**Research funding:** This work was funded by Research Ireland under grants 21/FFP-P/10187, 12/RC/2276_P2, 12/RC/2278_2 and 18/EPSRC-CDT/3585.

**Author contribution:** *All authors have accepted responsibility for the entire content of this manuscript and consented to its submission to the journal, reviewed all the results and approved the final version of the manuscript. A.L.B. conceived and coordinated the project. LF and CD assisted with the project. KM, CE and TF designed the experiments. HH developed the model code and performed the simulations with HK.*

**Conflict of interest**: Authors state no conflict of interest

**Data availability statement**: The datasets generated and/or analysed during the current study are available from the corresponding author upon reasonable request.

## References

[1] Y. S. Park, S. Guo, N. S. Makarov, and V. I. Klimov, "Room Temperature Single-Photon Emission from Individual Perovskite Quantum Dots," *ACS Nano,* vol. 9, no. 10, pp. 10386-93, Oct 27 2015, doi: 10.1021/acsnano.5b04584.

[2] P. Senellart, G. Solomon, and A. White, "High-performance semiconductor quantum-dot single-photon sources," *Nat Nanotechnol,* vol. 12, no. 11, pp. 1026-1039, Nov 7 2017, doi: 10.1038/nnano.2017.218.

[3] C. R. Kagan, E. Lifshitz, E. H. Sargent, and D. V. Talapin, "Building devices from colloidal quantum dots," *Science,* vol. 353, no. 6302, Aug 26 2016, doi: 10.1126/science.aac5523.

[4] A. Muravitskaya *et al.*, "Plasmon Mediated Long-Range Interaction Between Quantum Emitters and Metasurface: Angle-Controlled Unidirectional Emission Outcoupling," (in English), *Laser & Photonics Reviews,* Jul 22 2025, doi: 10.1002/lpor.202500656.

[5] H. Aouani, O. Mahboub, E. Devaux, H. Rigneault, T. W. Ebbesen, and J. Wenger, "Plasmonic antennas for directional sorting of fluorescence emission," *Nano Lett,* vol. 11, no. 6, pp. 2400-6, Jun 8 2011, doi: 10.1021/nl200772d.

[6] J. Guan *et al.*, "Engineering Directionality in Quantum Dot Shell Lasing Using Plasmonic Lattices," *Nano Lett,* vol. 20, no. 2, pp. 1468-1474, Feb 12 2020, doi: 10.1021/acs.nanolett.9b05342.

[7] E. Pelucchi *et al.*, "The potential and global outlook of integrated photonics for quantum technologies," *Nature Reviews Physics,* vol. 4, no. 3, pp. 194-208, 2021, doi: 10.1038/s42254-021-00398-z.

[8] A. Rossetti *et al.*, "Control and enhancement of optical nonlinearities in plasmonic semiconductor nanostructures," *Light Sci Appl,* vol. 14, no. 1, p. 192, May 13 2025, doi: 10.1038/s41377-025-01783-4.

[9] T. B. Hoang, G. M. Akselrod, and M. H. Mikkelsen, "Ultrafast Room-Temperature Single Photon Emission from Quantum Dots Coupled to Plasmonic Nanocavities," *Nano Lett,* vol. 16, no. 1, pp. 270-5, Jan 13 2016, doi: 10.1021/acs.nanolett.5b03724.

[10] S. Hu *et al.*, "Robust consistent single quantum dot strong coupling in plasmonic nanocavities," *Nat Commun,* vol. 15, no. 1, p. 6835, Aug 9 2024, doi: 10.1038/s41467-024-51170-7.

[11] X. Dai *et al.*, "Solution-processed, high-performance light-emitting diodes based on quantum dots," *Nature,* vol. 515, no. 7525, pp. 96-9, Nov 6 2014, doi: 10.1038/nature13829.

[12] L. Zhang, R. Gogna, W. Burg, E. Tutuc, and H. Deng, "Photonic-crystal exciton-polaritons in monolayer semiconductors," *Nat Commun,* vol. 9, no. 1, p. 713, Feb 19 2018, doi: 10.1038/s41467-018-03188-x.

[13] S. M. Thon *et al.*, "Strong coupling through optical positioning of a quantum dot in a photonic crystal cavity," *Applied Physics Letters,* vol. 94, no. 11, 2009, doi: 10.1063/1.3103885.

[14] P. M. Vora, A. S. Bracker, S. G. Carter, M. Kim, C. S. Kim, and D. Gammon, "Strong coupling of a quantum dot molecule to a photonic crystal cavity," *Physical Review B,* vol. 99, no. 16, 2019, doi: 10.1103/PhysRevB.99.165420.

[15] K. Kuruma, Y. Ota, M. Kakuda, S. Iwamoto, and Y. Arakawa, "Strong coupling between a single quantum dot and an L4/3 photonic crystal nanocavity," *Applied Physics Express,* vol. 13, no. 8, 2020, doi: 10.35848/1882-0786/aba7a8.

[16] J. Lawless, C. Hrelescu, C. Elliott, L. Peters, N. McEvoy, and A. L. Bradley, "Influence of Gold Nano-Bipyramid Dimensions on Strong Coupling with Excitons of Monolayer MoS(2)," *ACS Appl Mater Interfaces,* vol. 12, no. 41, pp. 46406-46415, Oct 14 2020, doi: 10.1021/acsami.0c09261.

[17] M. Stuhrenberg *et al.*, "Strong Light-Matter Coupling between Plasmons in Individual Gold Bi-pyramids and Excitons in Mono- and Multilayer WSe(2)," *Nano Lett,* vol. 18, no. 9, pp. 5938-5945, Sep 12 2018, doi: 10.1021/acs.nanolett.8b02652.

[18] R. Chikkaraddy *et al.*, "Single-molecule strong coupling at room temperature in plasmonic nanocavities," *Nature,* vol. 535, no. 7610, pp. 127-30, Jul 7 2016, doi: 10.1038/nature17974.

[19] R. Liu *et al.*, "Strong Light-Matter Interactions in Single Open Plasmonic Nanocavities at the Quantum Optics Limit," *Phys Rev Lett,* vol. 118, no. 23, p. 237401, Jun 9 2017, doi: 10.1103/PhysRevLett.118.237401.

[20] O. S. Ojambati *et al.*, "Quantum electrodynamics at room temperature coupling a single vibrating molecule with a plasmonic nanocavity," *Nat Commun,* vol. 10, no. 1, p. 1049, Mar 5 2019, doi: 10.1038/s41467-019-08611-5.

[21] J. Y. Li *et al.*, "Room-Temperature Strong Coupling Between a Single Quantum Dot and a Single Plasmonic Nanoparticle," *Nano Lett,* vol. 22, no. 12, pp. 4686-4693, Jun 22 2022, doi: 10.1021/acs.nanolett.2c00606.

[22] H. Gross, J. M. Hamm, T. Tufarelli, O. Hess, and B. Hecht, "Near-field strong coupling of single quantum dots," *Sci Adv,* vol. 4, no. 3, p. eaar4906, Mar 2018, doi: 10.1126/sciadv.aar4906.

[23] C. R. Kagan, L. C. Bassett, C. B. Murray, and S. M. Thompson, "Colloidal Quantum Dots as Platforms for Quantum Information Science," *Chem Rev,* vol. 121, no. 5, pp. 3186-3233, Mar 10 2021, doi: 10.1021/acs.chemrev.0c00831.

[24] M. Chen *et al.*, "Approaches for Positioning the Active Medium in Hybrid Nanoplasmonics. Focus on Plasmon-Assisted Photopolymerization," *ACS Photonics,* vol. 11, no. 10, pp. 3933-3953, Oct 16 2024, doi: 10.1021/acsphotonics.4c00868.

[25] A. Badolato *et al.*, "Deterministic coupling of single quantum dots to single nanocavity modes," *Science,* vol. 308, no. 5725, pp. 1158-61, May 20 2005, doi: 10.1126/science.1109815.

[26] S. H. Gong *et al.*, "Self-aligned deterministic coupling of single quantum emitter to nanofocused plasmonic modes," *Proc Natl Acad Sci U S A,* vol. 112, no. 17, pp. 5280-5, Apr 28 2015, doi: 10.1073/pnas.1418049112.

[27] S. Kim, S. H. Gong, J. H. Cho, and Y. H. Cho, "Unidirectional Emission of a Site-Controlled Single Quantum Dot from a Pyramidal Structure," *Nano Lett,* vol. 16, no. 10, pp. 6117-6123, Oct 12 2016, doi: 10.1021/acs.nanolett.6b02331.

[28] S.-H. Gong, S. Kim, J.-H. Kim, J.-H. Cho, and Y.-H. Cho, "Site-Selective, Two-Photon Plasmonic Nanofocusing on a Single Quantum Dot for Near-Room-Temperature Operation," *ACS Photonics,* vol. 5, no. 3, pp. 711-717, 2018, doi: 10.1021/acsphotonics.7b01238.

[29] N. Somaschi *et al.*, "Near-optimal single-photon sources in the solid state," *Nature Photonics,* vol. 10, no. 5, pp. 340-345, 2016, doi: 10.1038/nphoton.2016.23.

[30] D. Ratchford, F. Shafiei, S. Kim, S. K. Gray, and X. Li, "Manipulating coupling between a single semiconductor quantum dot and single gold nanoparticle," *Nano Lett,* vol. 11, no. 3, pp. 1049-54, Mar 9 2011, doi: 10.1021/nl103906f.

[31] H. Abudayyeh et al., "Single photon sources with near unity collection efficiencies by deterministic placement of quantum dots in nanoantennas," (in English), *Apl Photonics,* vol. 6, no. 3, Mar 1 2021, doi: Artn 03610910.1063/5.0034863.

[32] N. Nikolay *et al.*, "Accurate placement of single nanoparticles on opaque conductive structures," Applied Physics Letters, vol. 113, no. 11, 2018, doi: 10.1063/1.5049082.

[33] Y. Cui, M. T. Björk, J. A. Liddle, C. Sönnichsen, B. Boussert, and A. P. Alivisatos, "Integration of Colloidal Nanocrystals into Lithographically Patterned Devices," Nano Letters, vol. 4, no. 6, pp. 1093-1098, 2004, doi: 10.1021/nl049488i.

[34] K. Santhosh, O. Bitton, L. Chuntonov, and G. Haran, "Vacuum Rabi splitting in a plasmonic cavity at the single quantum emitter limit," *Nat Commun,* vol. 7, p. ncomms11823, Jun 13 2016, doi: 10.1038/ncomms11823.

[35] C. Ropp, Z. Cummins, S. Nah, J. T. Fourkas, B. Shapiro, and E. Waks, "Nanoscale imaging and spontaneous emission control with a single nano-positioned quantum dot," *Nat Commun,* vol. 4, p. 1447, 2013, doi: 10.1038/ncomms2477.

[36] S. Ghosh and A. Ghosh, "All optical dynamic nanomanipulation with active colloidal tweezers," *Nat Commun,* vol. 10, no. 1, p. 4191, Sep 13 2019, doi: 10.1038/s41467-019-12217-2.

[37] P. Vasa and C. Lienau, "Strong Light–Matter Interaction in Quantum Emitter/Metal Hybrid Nanostructures," *ACS Photonics,* vol. 5, no. 1, pp. 2-23, 2017, doi: 10.1021/acsphotonics.7b00650.

[38] M. Hertzog, M. Wang, J. Mony, and K. Borjesson, "Strong light-matter interactions: a new direction within chemistry," *Chem Soc Rev,* vol. 48, no. 3, pp. 937-961, Feb 4 2019, doi: 10.1039/c8cs00193f.

[39] D. J. Tibben, G. O. Bonin, I. Cho, G. Lakhwani, J. Hutchison, and D. E. Gomez, "Molecular Energy Transfer under the Strong Light-Matter Interaction Regime," *Chem Rev,* vol. 123, no. 13, pp. 8044-8068, Jul 12 2023, doi: 10.1021/acs.chemrev.2c00702.

[40] G. M. Akselrod *et al.*, "Probing the mechanisms of large Purcell enhancement in plasmonic nanoantennas," *Nature Photonics,* vol. 8, no. 11, pp. 835-840, 2014, doi: 10.1038/nphoton.2014.228.

[41] D. Ge *et al.*, "Hybrid plasmonic nano-emitters with controlled single quantum emitter positioning on the local excitation field," *Nat Commun,* vol. 11, no. 1, p. 3414, Jul 8 2020, doi: 10.1038/s41467-020-17248-8.

[42] S. Morozov *et al.*, "Electrical control of single-photon emission in highly charged individual colloidal quantum dots," *Sci Adv,* vol. 6, no. 38, Sep 2020, doi: 10.1126/sciadv.abb1821.

[43] J. Huang *et al.*, "Plasmon-Induced Trap State Emission from Single Quantum Dots," *Phys Rev Lett,* vol. 126, no. 4, p. 047402, Jan 29 2021, doi: 10.1103/PhysRevLett.126.047402.

[44] O. Iff *et al.*, "Purcell-Enhanced Single Photon Source Based on a Deterministically Placed WSe(2) Monolayer Quantum Dot in a Circular Bragg Grating Cavity," *Nano Lett,* vol. 21, no. 11, pp. 4715-4720, Jun 9 2021, doi: 10.1021/acs.nanolett.1c00978.

[45] J. Huang, A. J. Traverso, G. Yang, and M. H. Mikkelsen, "Real-Time Tunable Strong Coupling: From Individual Nanocavities to Metasurfaces," *ACS Photonics,* vol. 6, no. 4, pp. 838-843, 2019, doi: 10.1021/acsphotonics.8b01743.

[46] Y. Chen *et al.*, "Metasurface Integrated Monolayer Exciton Polariton," *Nano Lett,* vol. 20, no. 7, pp. 5292-5300, Jul 8 2020, doi: 10.1021/acs.nanolett.0c01624.

[47] X. Zhang and A. L. Bradley, "Polaritonic critical coupling in a hybrid quasibound states in the continuum cavity–WS2 monolayer system," Physical Review B, vol. 105, no. 16, 2022, doi: 10.1103/PhysRevB.105.165424.

[48] T. Weber et al., "Intrinsic strong light-matter coupling with self-hybridized bound states in the continuum in van der Waals metasurfaces," Nat Mater, vol. 22, no. 8, pp. 970-976, Aug 2023, doi: 10.1038/s41563-023-01580-7.

[49] H. Hajian *et al.*, "Quasi-bound states in the continuum for electromagnetic induced transparency and strong excitonic coupling," *Opt Express,* vol. 32, no. 11, pp. 19163-19174, May 20 2024, doi: 10.1364/OE.525535.

[50] T. T. H. Do *et al.*, "Room-temperature strong coupling in a single-photon emitter-metasurface system," *Nat Commun,* vol. 15, no. 1, p. 2281, Mar 13 2024, doi: 10.1038/s41467-024-46544-w.

[51] X. Xiong *et al.*, "Control of Plexcitonic Strong Coupling via Substrate-Mediated Hotspot Nanoengineering," *Advanced Optical Materials,* vol. 10, no. 17, 2022, doi: 10.1002/adom.202200557.

[52] H. Leng, B. Szychowski, M. C. Daniel, and M. Pelton, "Strong coupling and induced transparency at room temperature with single quantum dots and gap plasmons," *Nat Commun,* vol. 9, no. 1, p. 4012, Oct 1 2018, doi: 10.1038/s41467-018-06450-4.

[53] K. D. Park *et al.*, "Tip-enhanced strong coupling spectroscopy, imaging, and control of a single quantum emitter," *Sci Adv,* vol. 5, no. 7, p. eaav5931, Jul 2019, doi: 10.1126/sciadv.aav5931.

[54] K. Ueno *et al.*, "Nanoparticle plasmon-assisted two-photon polymerization induced by incoherent excitation source," *J Am Chem Soc,* vol. 130, no. 22, pp. 6928-9, Jun 4 2008, doi: 10.1021/ja801262r.

[55] X. Zhou *et al.*, "Selective Functionalization of the Nanogap of a Plasmonic Dimer," *ACS Photonics,* vol. 2, no. 1, pp. 121-129, 2014, doi: 10.1021/ph500331c.

[56] X. Zhou *et al.*, "Two-Color Single Hybrid Plasmonic Nanoemitters with Real Time Switchable Dominant Emission Wavelength," *Nano Lett,* vol. 15, no. 11, pp. 7458-66, Nov 11 2015, doi: 10.1021/acs.nanolett.5b02962.

[57] Y. M. Morozov *et al.*, "Plasmon-Enhanced Multiphoton Polymer Crosslinking for Selective Modification of Plasmonic Hotspots," *J Phys Chem C Nanomater Interfaces,* vol. 128, no. 43, pp. 18641-18650, Oct 31 2024, doi: 10.1021/acs.jpcc.4c05936.

[58] M. Liu and P. Guyot-Sionnest, "Mechanism of silver(I)-assisted growth of gold nanorods and bipyramids," *J Phys Chem B,* vol. 109, no. 47, pp. 22192-200, Dec 1 2005, doi: 10.1021/jp054808n.

[59] N. R. Jana, L. Gearheart, and C. J. Murphy, "Seed-Mediated Growth Approach for Shape-Controlled Synthesis of Spheroidal and Rod-like Gold Nanoparticles Using a Surfactant Template," *Advanced Materials,* vol. 13, no. 18, pp. 1389-1393, 2001, doi: 10.1002/1521-4095(200109)13:18<1389::Aid-adma1389>3.0.Co;2-f.

[60] "Lumerical Inc." www.lumerical.com/tcadproducts/fdtd (accessed 2025.

[61] P. B. Johnson and R. W. Christy, "Optical Constants of the Noble Metals," *Physical Review B,* vol. 6, no. 12, pp. 4370-4379, 1972, doi: 10.1103/PhysRevB.6.4370.

**Supplementary Material:** This article contains supplementary material (https://doi.org/---). Additional details of the materials, employed approaches and simulations results for this study